\documentclass[aps,twocolumn,preprintnumbers,amsmath,amssymb,superscriptaddress,showpacs]{revtex4-2}

\usepackage{graphicx}
\usepackage{amsmath}
\usepackage{multirow}
\usepackage{xcolor}
\usepackage{ulem}
\usepackage[version=4]{mhchem}
\newcommand{\RNum}[1]{\uppercase\expandafter{\romannumeral #1\relax}}

\begin{document}

\title{Symmetry-Preserving Phase Transitions in $AM_2$Al$_9$ Materials under Pressure}

\author{Jian-Feng Zhang}\email{jianfeng.zhang@hpstar.ac.cn}
\affiliation{ Center for High Pressure Science and Technology Advanced Research, Beijing 100193, China. }
\author{Sheng Xu}\email{shengxu@zjut.edu.cn}
\affiliation{ School of Physics, Zhejiang University of Technology, Hangzhou 310023, China }
\affiliation{ Interdisciplinary Center for Quantum Information \& State Key Laboratory of Silicon and Advanced Semiconductor Materials, School of Physics, Zhejiang University, Hangzhou 310027, China }
\author{Zhong-Yi Lu}\email{zlu@ruc.edu.cn} 
\affiliation{ School of Physics and Beijing Key Laboratory of Opto-electronic Functional Materials \& Micro-nano Devices, Renmin University of China, Beijing 100872, China }
\affiliation{ Key Laboratory of Quantum State Construction and Manipulation (Ministry of Education), Renmin University of China, Beijing 100872, China }
\author{Tao Xiang}\email{txiang@iphy.ac.cn}
\affiliation{ Institute of Physics, Chinese Academy of Sciences, Beijing 100190, China}
\affiliation{ School of Physical Sciences, University of Chinese Academy of Sciences, Beijing 100049, China}

\date{\today}

\begin{abstract}

External parameters such as temperature, pressure, and chemical doping can induce structural phase transitions in materials. Although such transitions usually involve a change in symmetry, an uncommon exception is the isostructural phase transition, which is first order yet preserves the symmetry of the parent structure. Using first-principles calculations, we show that $AM_2$Al$_9$ compounds ($A$ = Ba, Ca, Sr, or Eu; $M$ = Fe, Co, or Ni) undergo pressure-induced isostructural phase transitions. At the transition pressure, these systems exhibit a pronounced volume collapse while retaining the same crystal symmetry and space group ($P6/mmm$). Bonding analysis based on the integrated crystal orbital Hamilton population (ICOHP) shows that the transition is driven by a redistribution of bonding character between intralayer and interlayer atomic bonds. Because isostructural transitions are rare in single crystals, $AM_2$Al$_9$ provides a promising platform for investigating critical phenomena under pressure and for deepening our understanding of symmetry-preserving structural transitions.

\end{abstract}


\maketitle

\section{INTRODUCTION}

 In crystalline functional materials, atomic structure and crystal symmetry are fundamental to understanding physical properties. Temperature, pressure, doping, and other external parameters can drive structural phase transitions and reveal emergent phenomena. According to the continuity of the Gibbs free energy, phase transitions are conventionally classified as first or second order. Structural transitions of either type are often identified by a change in crystal symmetry or space group.

 For a second-order structural transition, Landau theory requires the spontaneous breaking of a continuous symmetry, so the high- and low-symmetry phases must have a group--subgroup relationship. Representative examples include charge-density-wave (CDW) transitions driven by strong electron--phonon coupling (EPC), which break translational symmetry~\cite{NbSe2,TaS2}, and ferroelectric (FE) transitions, which break inversion symmetry~\cite{BaTiO3,PbTiO3}. First-order structural transitions, by contrast, are not required to change symmetry. A material may undergo a discontinuous structural transformation while retaining its original symmetry; such a transformation is known as an isostructural phase transition~\cite{Ce}.

 Isostructural phase transitions have been reported in only a small number of single-crystalline solids~\cite{Ce,VO2,WSe2,A2Ir2O7,SmS}. The best-known example is the $\alpha$--$\gamma$ transition in elemental Ce~\cite{Ce}, which is driven by strongly correlated $4f$ electrons~\cite{Ce-cal,Ce-cp}. Pressure-induced isostructural transitions have also been predicted in Pr, Ca, Fe, and Zr~\cite{Pr,Ca,Fe,Zr}. Such transitions can accompany other unusual phenomena, including enhanced superconductivity in Sn$_4$P$_3$~\cite{Sn4P3} and insulator-to-metal transitions in TiS$_3$ and WSe$_2$~\cite{TiS3,WSe2}. A distinctive feature of an isostructural transition is the possible existence of a critical point, above which the two symmetry-equivalent phases merge continuously. Critical-point behavior is well established in fluids such as water, carbon dioxide, and nitrogen, but the scarcity of isostructural transitions in single crystals has made analogous critical points in solids difficult to identify~\cite{Ce-cp,BFO-PTO,SmS}.

 The $AM_2$Al$_9$ system ($A$ = Ba, Ca, Sr, and Eu; $M$ = Fe, Co, and Ni) was first synthesized in the 1980s~\cite{129-1,129-2,129-3,129-4}. 
 Its complex lattice combines triangular, honeycomb, and kagome motifs and hosts phenomena such as frustrated magnetism, Dirac-like dispersions, and kagome flat bands~\cite{BaCoAl}. 
 Recently, BaFe$_2$Al$_9$ has attracted considerable attention because of its anomalous first-order CDW transition~\cite{BaFeAl-1,BaFeAl-2,BaFeAl-3,BaFeAl-4}. 
 Although the mechanism remains under debate, it is believed to be related to the partially filled Fe $3d$ orbitals. 
 More recently, the EuCo$_2$Al$_9$ system was reported to exhibit a giant anomalous Hall effect (AHE), which is believed to originate from its noncoplanar spin textures~\cite{EuCoAl,xu}. 
 Despite this growing interest, the effects of pressure on the crystal and electronic structures of $AM_2$Al$_9$ remain largely unexplored. 
 
 Here, using first-principles calculations, we demonstrate that this material family undergoes pressure-induced isostructural phase transitions. Section II describes the computational and bonding-analysis methods. Section III outlines the defining features of isostructural transitions. Section IV presents the calculated transitions in $AM_2$Al$_9$, Section V discusses the potential critical point, and Section VI identifies the microscopic bonding mechanism. Section VII summarizes the main conclusions.

\section{Computational Methods}

 The total energies and structural properties of $AM_2$Al$_9$ ($A$ = Ba, Ca, Sr, or Eu; $M$ = Fe, Co, or Ni) under pressure were investigated using density functional theory (DFT)~\cite{dft1,dft2} as implemented in the Vienna \textit{Ab initio} Simulation Package (VASP)~\cite{vasp1,vasp2}. 
 In total, ten materials were studied, including EuCo$_2$Al$_9$, CaFe$_2$Al$_9$, CaCo$_2$Al$_9$, CaNi$_2$Al$_9$, SrFe$_2$Al$_9$, SrCo$_2$Al$_9$, SrNi$_2$Al$_9$, BaFe$_2$Al$_9$, BaCo$_2$Al$_9$, and BaNi$_2$Al$_9$. 
 Exchange and correlation were treated within the Perdew--Burke--Ernzerhof generalized-gradient approximation~\cite{pbe}. The plane-wave kinetic-energy cutoff was 350 eV, and the Brillouin zone was sampled using an $8\times8\times12$ $\mathbf{k}$-point mesh. Gaussian smearing with a width of 0.05 eV was used for Fermi-surface broadening, and spin--orbit coupling (SOC) was included. During structural relaxation, both the lattice parameters and internal atomic coordinates were optimized until the residual force on every atom was below 0.01 eV/\AA. To examine possible correlation-induced magnetism associated with the Fe, Co, and Ni $3d$ orbitals, we also performed GGA+$U$ calculations with an effective Hubbard parameter $U=4$ eV~\cite{ldau}. Dynamical stability under pressure was evaluated using density-functional perturbation theory (DFPT)~\cite{dfptreview,dfptreview2} as implemented in Quantum ESPRESSO~\cite{pwscf}, with a $6\times6\times6$ $\mathbf{q}$-point mesh.

 Atomic-orbital projections and chemical-bonding properties were analyzed using maximally localized Wannier functions (MLWFs)~\cite{mlwf}. Orthogonal, atomic-orbital-like Wannier functions were used in place of nonorthogonal atomic orbitals. Bond strengths were quantified using the integrated crystal orbital Hamilton population (ICOHP)~\cite{cohp,cohp2}. In the MLWF basis, the ICOHP between atoms $A$(\textbf{0}) and $B$(\textbf{R}) is defined as
 \begin{equation}
 \label{eq_icohp}
     \text{ICOHP}_{AB}(\textbf{R}) = \sum_{\alpha\in A,\beta\in B}H_{A\alpha,B\beta}(\textbf{R})D^*_{A\alpha,B\beta}(\textbf{R})
 \end{equation}
 Here $A$ and $B$ label atoms in a unit cell, \textbf{R} is a real-space lattice vector, and $\alpha$ and $\beta$ label atomic orbitals on $A$ and $B$, respectively. $H_{A\alpha,B\beta}(\textbf{R})$ is the hopping integral and $D_{A\alpha,B\beta}(\textbf{R})$ is the corresponding reduced-density-matrix element (bond order).

\begin{figure*}[t]
\includegraphics[angle=0,scale=0.5]{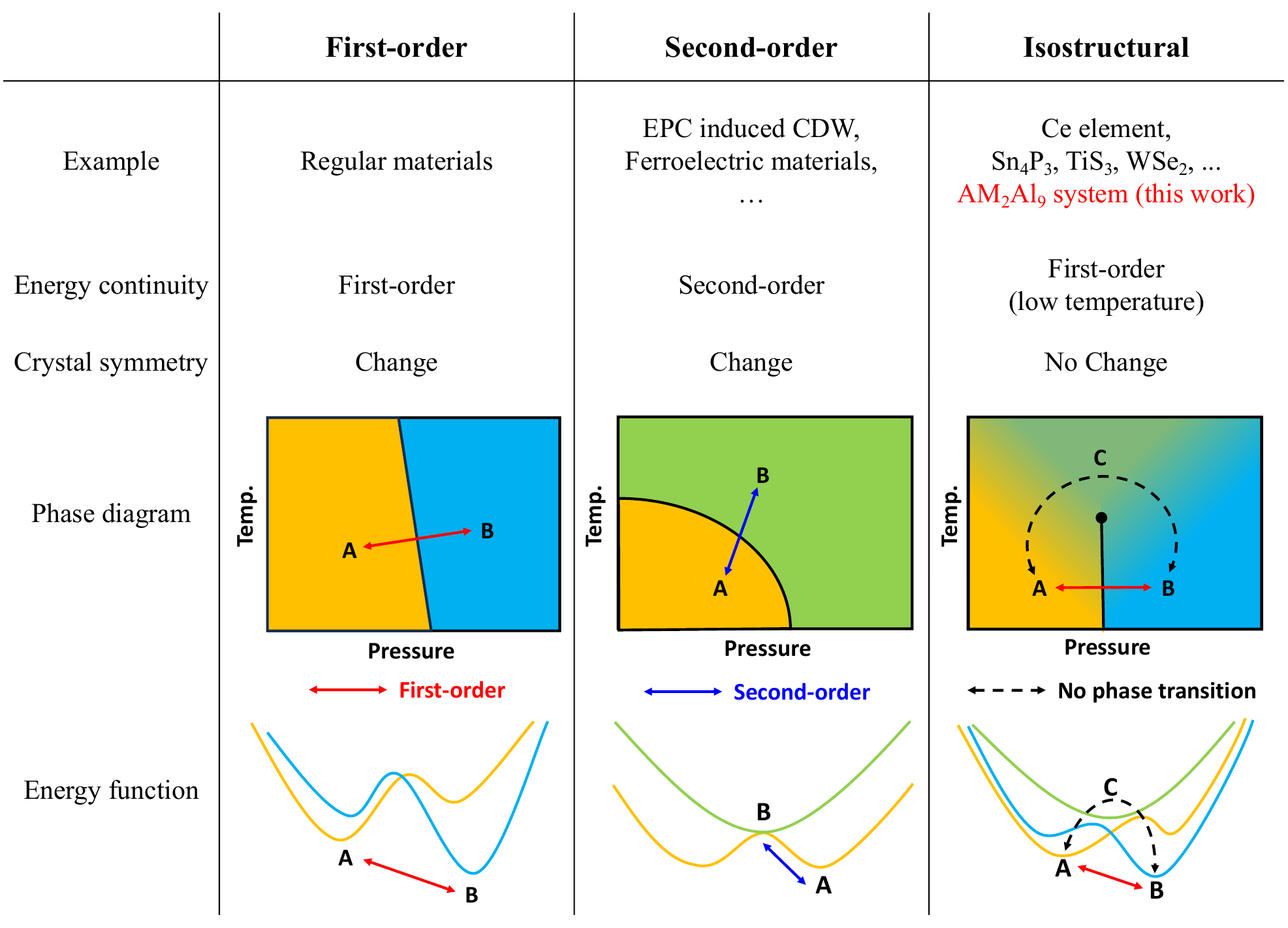}
\caption{
 Schematic comparison of symmetry-changing first-order, second-order, and isostructural phase transitions in single crystals. The upper panels show representative phase diagrams, and the lower panels show Gibbs-free-energy landscapes in structural-coordinate space. Red and blue arrows denote first- and second-order transitions between phases A and B, respectively. In the isostructural case, the black dashed arrow denotes continuous evolution above the critical point and therefore does not represent a phase transition.
 }
\label{Fig_Sum}
\end{figure*}

\section{Isostructural Phase Transitions}

 Figure \ref{Fig_Sum} compares symmetry-changing first- and second-order structural transitions with an isostructural transition. 
 The first two both involve changes in crystal symmetry, whereas the latter does not. 
 For symmetry-changing transitions, an order parameter can usually be defined directly from the broken symmetry within the Landau framework. Defining an analogous order parameter for a symmetry-preserving isostructural transition is more subtle~\cite{1982,1989}. Here we focus on transitions of the atomic structure and exclude electronic transitions, such as magnetic ordering, that themselves break symmetry and can also produce discontinuous lattice changes~\cite{Ce-cal}. Throughout the following discussion, the electronic ground state is assumed to be nonmagnetic and to share the symmetry of the atomic structure.

 The upper panels show representative pressure--temperature phase diagrams. The red and blue arrows indicate first- and second-order structural transitions between two distinct phases, respectively. The lower panels sketch the corresponding Gibbs-free-energy landscapes.
 For a first-order structural transition, the effect of pressure can be viewed as adjusting the relative positions of different energy wells. 
 The ground structural phase is determined by the lowest energy well. 
 In a second-order structural transition, atoms move continuously away from high-symmetry sites, and the crystal enters a lower-symmetry phase. In the free-energy landscape, a single high-symmetry minimum splits into symmetry-related low-symmetry minima.

 An isostructural transition has a qualitatively different phase diagram [right column of Fig. \ref{Fig_Sum}]. 
 At low temperatures, this transition is characterized by a symmetry-preserving first-order structural transition, as indicated by the red arrow. 
 At higher temperature, thermal fluctuations can remove the free-energy barrier and terminate the first-order line at a critical point. Above that point, phases A and B can evolve continuously into one another through an intermediate state C. Along this path (black dashed arrow), the Gibbs free energy remains analytic and the symmetry is unchanged; the path therefore does not cross a phase boundary.

 In fact, critical point behavior has been well-documented in many common substances like water, carbon dioxide, or nitrogen. 
 In these substances, the critical point marks the merging of two disordered phases, typically a liquid and a gas. In high-symmetry single-crystalline solids, by contrast, symmetry-preserving isostructural transitions have been reported in only a few materials~\cite{Ce,VO2,WSe2,A2Ir2O7,SmS}. 
 This rarity makes detecting a critical point in single-crystal materials particularly challenging~\cite{Ce-cp,BFO-PTO,SmS}. 
 We show that a family of $AM_2$Al$_9$ materials~\cite{129-1} undergoes such transitions under pressure. Because the calculated transition barrier is small, this family may provide an experimentally accessible platform for studying criticality in crystalline solids.

\section{Pressure-Induced Transitions in $AM_2$Al$_9$}

\begin{figure}[t]
\includegraphics[angle=0,scale=0.35]{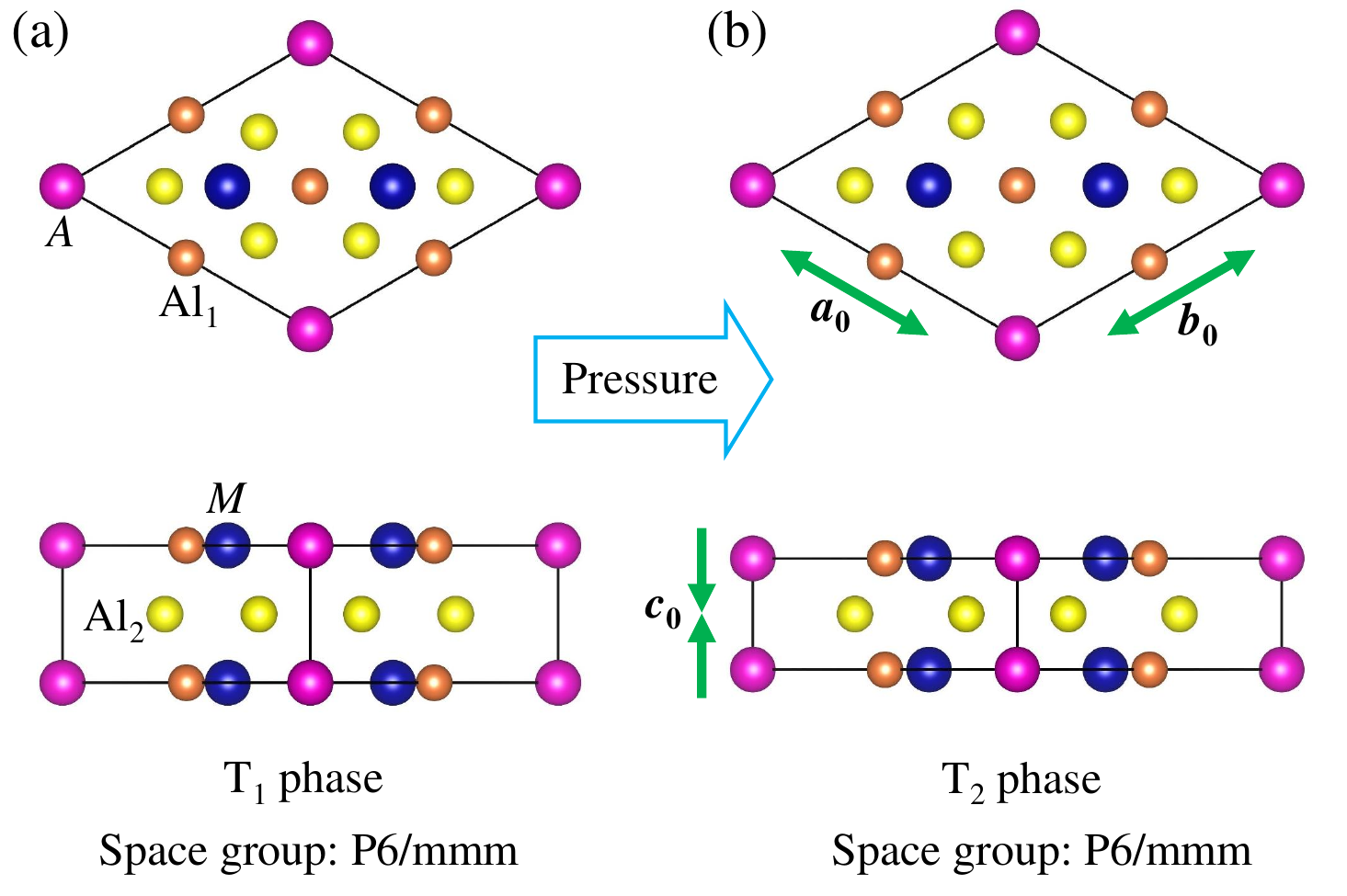}
\caption{
Top and side views of $AM_2$Al$_9$ in the (a) T$_1$ and (b) T$_2$ phases. Here $A$ = Ba, Ca, Sr, or Eu (purple); $M$ = Fe, Co, or Ni (blue); and Al$_1$ and Al$_2$ denote the two inequivalent Al sites (orange and yellow, respectively). Green arrows indicate the expansion of $a_0$ and $b_0$ and the collapse of $c_0$ under pressure.
 }
\label{Fig_str}
\end{figure}

 The crystal structure of the $AM_2$Al$_9$ ($A$ = Ba, Ca, Sr, and Eu, $M$ = Fe, Co, and Ni) materials is depicted in Fig. \ref{Fig_str}. 
 The materials crystallize in the P6/mmm space group~\cite{129-1}. 
 In this structure, the \textit{A} (purple) atoms form a triangular lattice, while the \textit{M} (blue) atoms are arranged in a honeycomb lattice. 
 There are two types of Al atoms: Al$_1$ (orange) atoms are arranged in a Kagome lattice, and Al$_2$ (yellow) atoms closely encircle the $M$ atoms. 
 The $A$, $M$, and Al$_1$ atoms lie in the same layer, whereas Al$_2$ occupies an intermediate layer. Under pressure, the lattice undergoes pronounced anisotropic flattening. 
 Its in-plane lattice parameters $\textbf{\textit{a}}_0$ and $\textit{\textbf{b}}_0$ expand abruptly, while the $\textbf{\textit{c}}_0$ parameter contracts sharply, as indicated by the green arrows in Figure \ref{Fig_str}(b). 
 Pressure therefore transforms the primitive trigonal lattice from the T$_1$ phase into a more strongly flattened T$_2$ phase. 
 Importantly, throughout this flattening, the crystal maintains its original symmetry under the space group P6/mmm. 
 
\begin{figure}[t]
\includegraphics[angle=0,scale=0.46]{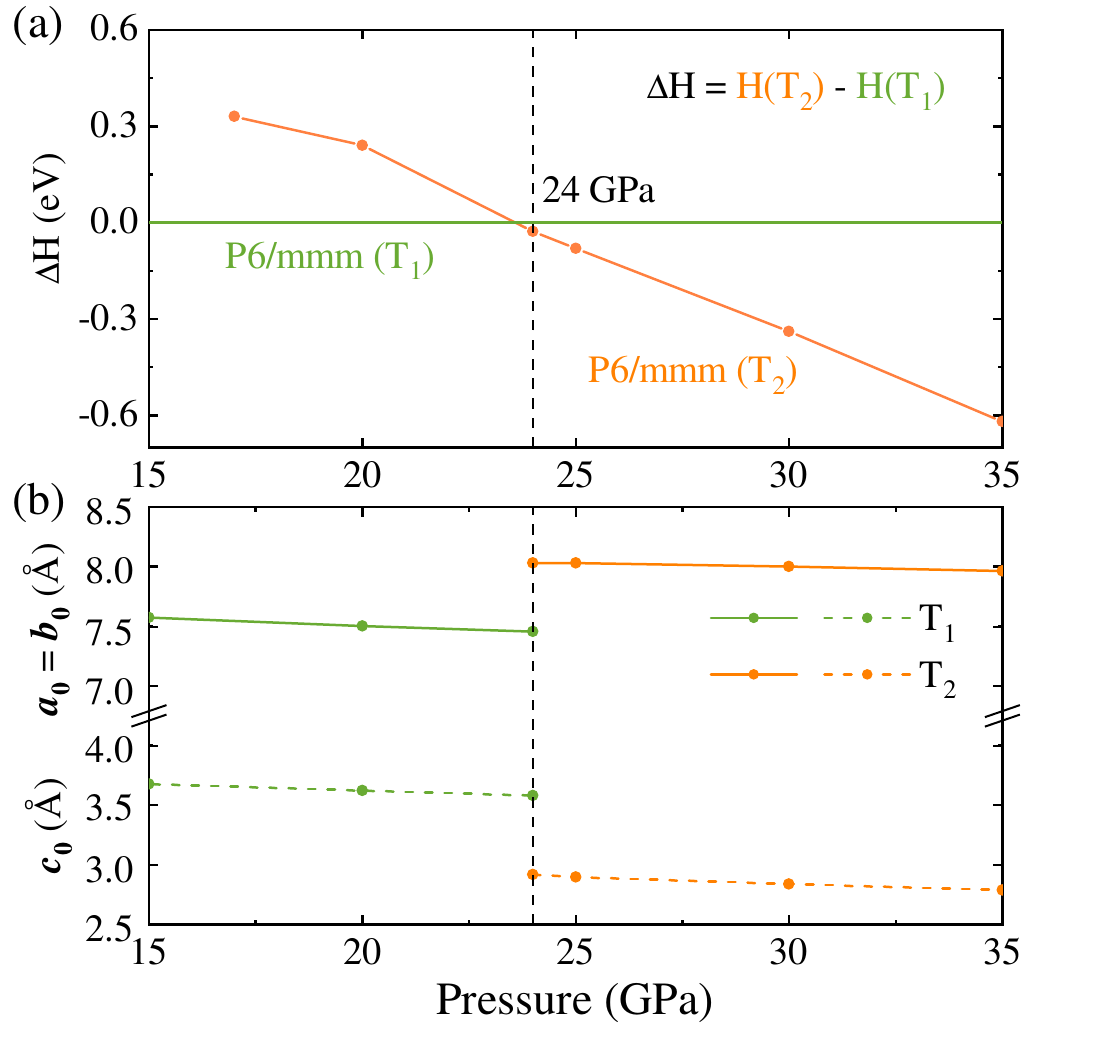}
\caption{
(a) Enthalpy difference $\Delta H=H(\mathrm{T}_2)-H(\mathrm{T}_1)$ for EuCo$_2$Al$_9$ near the transition pressure. (b) Pressure dependence of $a_0$ (solid curves) and $c_0$ (dashed curves) in the T$_1$ (green) and T$_2$ (orange) phases.
 }
\label{Fig_ent}
\end{figure}

 In addition to discontinuous changes in the lattice parameters and volume, the continuity of the Gibbs free energy is central to classifying a phase transition. 
 For instance, consider the material EuCo$_2$Al$_9$~\cite{EuCoAl,xu}. 
 Figure \ref{Fig_ent}(a) displays the enthalpy difference between its T$_1$ and T$_2$ phases under pressure: $\Delta H = H(\text{T}_2) - H(\text{T}_1)$. 
 An intersection of their enthalpy curves around 24 GPa indicates a typical pressure-induced first-order phase transition. 
 Figure \ref{Fig_ent}(b) shows the accompanying discontinuous changes in the lattice parameters. 
 Near the transition pressure of 24 GPa, the in-plane lattice parameters \textit{\textbf{a$_0$}} and \textit{\textbf{b$_0$}} (solid lines) expand by about 8\%, while the \textit{\textbf{c$_0$}} lattice parameter (dash lines) contracts by about 19\%. In total, the volume of unit cell decreases by about 5\%.

 The phonon structures of EuCo$_2$Al$_9$ under various pressures and structural phases were calculated (See Fig. S2 of Supplemental materials). 
 The absence of imaginary frequencies confirms that both the T$_1$ and T$_2$ phases are dynamically stable and argues against a strong EPC-driven CDW instability. 
 We also examined possible correlation-induced magnetism on the Co atoms. 
 Our calculations show that the Co atoms do not exhibit any local magnetic moments, even when an effective Hubbard U of 4 eV is applied on their 3d orbitals. 
 This outcome is expected since the Co atomic 3d orbitals are nearly fully filled. 
 The absence of charge and magnetic ordering indicates that the T$_1$--T$_2$ transition is not driven by electronic symmetry breaking. 
 The symmetry of the electronic ground state always remains consistent with the crystal symmetry throughout the transition. 
 Together with the discontinuous lattice and enthalpy changes, the identical symmetries of T$_1$ and T$_2$ establish the transition as isostructural. 

 All ten compounds examined exhibit the same qualitative isostructural transition. 
 The transition pressures and corresponding lattice flattening are detailed in TABLE \ref{table:lattice}. 
 Compounds with $A$ = Ba have the highest transition pressures, whereas those with $A$ = Ca have the lowest. 
 This can be attributed to the larger atomic radius of the Ba atom, which effectively provides a negative chemical pressure. 
 Moreover, systems with $M$ = Co and Ni generally display similar transition pressures, which are significantly higher than those in systems with $M$ = Fe. 
 This trend recalls the unusual first-order CDW transition in BaFe$_2$Al$_9$. Although its microscopic origin remains unclear, it is thought to involve the partially filled Fe $3d$ orbitals. These Fe valence states may also contribute to the lower transition pressures of the Fe-based compounds, as discussed below. 

\begin{table*}[tb]
\caption{\label{table:lattice}
Calculated transition pressures and lattice parameters of $AM_2$Al$_9$. The lattice parameters are given in \AA.}
\begin{center}
\begin{tabular*}{16cm}{@{\extracolsep{\fill}} cccccc}

\hline \hline
Material & Transition pressure & \textbf{\textit{a$_0$}} (T$_1$) & \textbf{\textit{c$_0$}} (T$_1$) & \textbf{\textit{a$_0$}} (T$_2$) & \textbf{\textit{c$_0$}} (T$_2$)  \\
\hline
EuCo$_2$Al$_9$ & 24.1 GPa & 7.454 & 3.583 & 8.032 & 2.918 \\
CaFe$_2$Al$_9$ & 5.6 GPa & 7.971 & 3.547 & 8.245 & 3.176 \\
CaCo$_2$Al$_9$ & 17.1 GPa & 7.568 & 3.606 & 7.992 & 3.090 \\
CaNi$_2$Al$_9$ & 16.1 GPa & 7.599 & 3.626 & 8.272 & 2.875 \\
SrFe$_2$Al$_9$ & 15.4 GPa & 7.767 & 3.528 & 8.140 & 3.083 \\
SrCo$_2$Al$_9$ & 31.3 GPa & 7.388 & 3.550 & 7.925 & 2.943 \\
SrNi$_2$Al$_9$ & 28.1 GPa & 7.440 & 3.576 & 8.171 & 2.802 \\
BaFe$_2$Al$_9$ & 26.6 GPa & 7.572 & 3.543 & 8.015 & 3.046 \\
BaCo$_2$Al$_9$ & 47.3 GPa & 7.239 & 3.509 & 7.861 & 2.837 \\
BaNi$_2$Al$_9$ & 42.5 GPa & 7.301 & 3.522 & 8.012 & 2.792 \\
\hline\hline
\end{tabular*}
\end{center}
\end{table*}

 \section{Critical Point in $AM_2$Al$_9$}

 A defining feature of an isostructural phase transition is the possible termination of the first-order line at a critical point. 
 As depicted in the third column of Fig. \ref{Fig_Sum}, the symmetry-equivalent T$_1$ and T$_2$ phases of $AM_2$Al$_9$ are represented by two minima in the enthalpy landscape (yellow and blue curves, respectively). 
 At low temperatures, the pressure-induced transition between T$_1$ and T$_2$ (indicated by the red arrow) remains first-order. 
 At higher temperatures, thermal fluctuations of the electrons and nuclei can wash out the Gibbs-free-energy barrier, allowing the T$_1$ and T$_2$ phases to merge continuously into a single state and terminating the first-order boundary at a critical point. 
 This temperature effect is analogous to that in temperature-dependent second-order structural transitions, such as the strong EPC induced CDW or the FE transitions. 
 In these cases, the energy barrier corresponds to the energy gain from their respective symmetry breakings: periodic translational symmetry for CDW materials and inversion symmetry for FE materials. 
 The temperature at which this barrier disappears coincides with the CDW or FE transition temperature, where the broken symmetry is restored. 
 This restoration of symmetry also marks a significant difference between second-order and isostructural phase transitions. 
 In the latter, all transition pathways, as illustrated by both the red arrow and black dashed arrow of Fig. \ref{Fig_Sum}, do not involve any changes in symmetry. 

\begin{figure}[t]
\includegraphics[angle=0,scale=0.46]{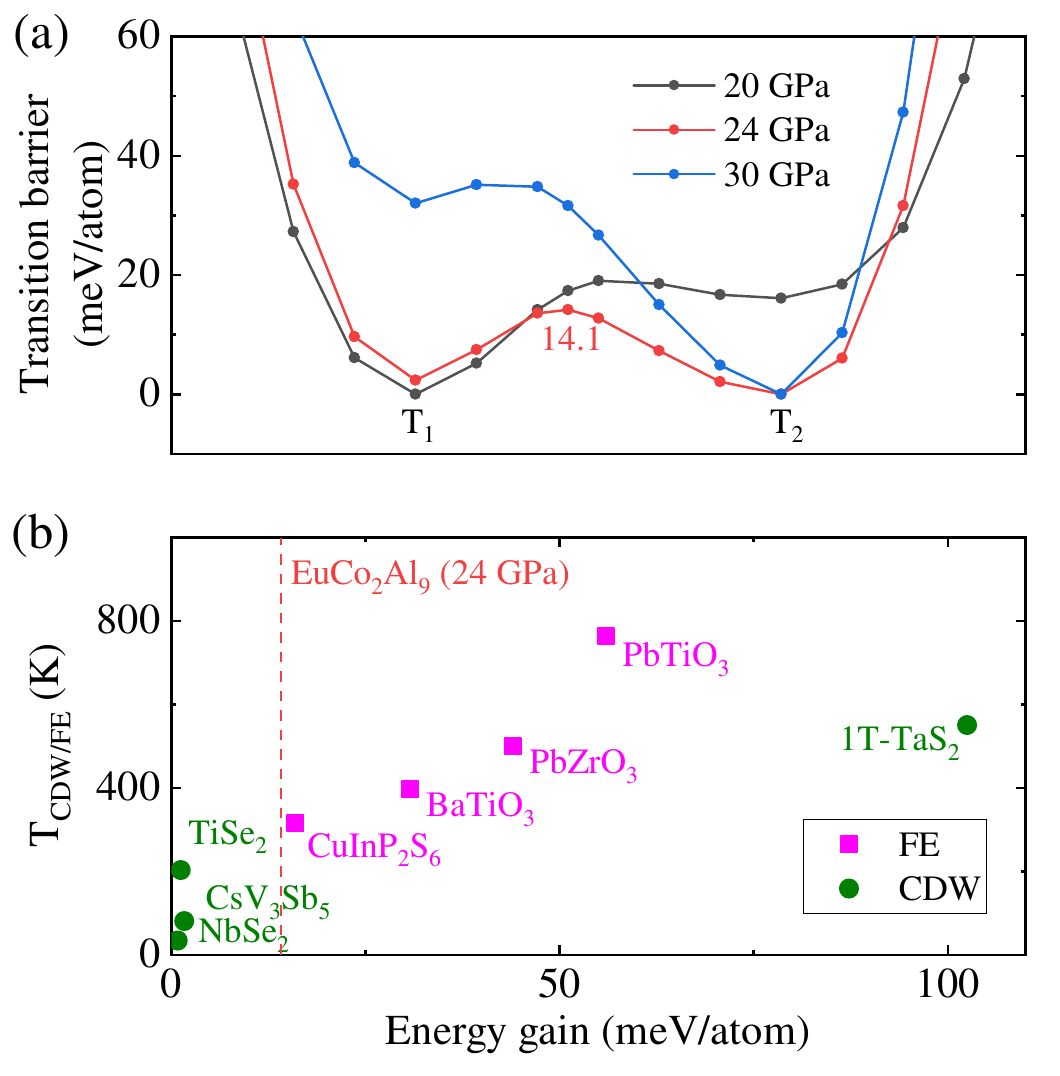}
\caption{
(a) Enthalpy profile along a linear interpolation from T$_1$ to T$_2$ in EuCo$_2$Al$_9$ at 20, 24, and 30 GPa. (b) Comparison with characteristic energy gains and transition temperatures in representative second-order FE materials [BaTiO$_3$~\cite{BaTiO3-cal}, PbZrO$_3$~\cite{PbZrO3-cal}, PbTiO$_3$~\cite{PbTiO3-cal}, and CuInP$_2$S$_6$~\cite{CuInP2S6-cal}] and CDW materials [2H-NbSe$_2$~\cite{NbSe2-cal}, 1T-TaS$_2$~\cite{TaS2-cal}, 1T-TiSe$_2$~\cite{TiSe2-cal}, and CsV$_3$Sb$_5$~\cite{CsV3Sb5-cal}].
 }
\label{Fig_barr}
\end{figure}

 Experimental access to such a critical point is favored by a small transition barrier and therefore a relatively low critical temperature. 
 Typically, the energy barrier for a first-order structural transition is significantly higher than that for a second-order transition. 
 However, in the case of the first-order T$_1$-T$_2$ transition in $AM_2$Al$_9$ materials, the barrier is naturally smaller due to their identical crystal symmetry and closely packed atomic distributions.
 For EuCo$_2$Al$_9$, we estimated the barrier by linearly interpolating between the T$_1$ and T$_2$ structures, as shown in Fig. \ref{Fig_barr}(a). 
 Near its transition pressure of 24 GPa (indicated by the red line), the enthalpy barrier reaches its minimum of about 14.1 meV/atom.

Figure \ref{Fig_barr}(b) displays several materials that undergo second-order FE (purple) or CDW (green) transitions. 
We have plotted their characteristic transition temperatures (T$_\text{CDW}$ or T$_\text{FE}$) alongside the energy gains associated with their respective symmetry breakings. 
The calculated barrier of 14.1 meV/atom is comparable to, or smaller than, the characteristic energy scales of several FE and CDW materials. Thermal fluctuations at temperatures of a few hundred kelvin could therefore plausibly overcome it. Above the resulting critical temperature, the T$_1$ and T$_2$ structures would merge continuously while preserving their common symmetry. These results make $AM_2$Al$_9$ a promising platform for exploring critical-point behavior in a single-crystalline solid.

\section{Microscopic Origin of the Transition}

\begin{figure}[t]
\includegraphics[angle=0,scale=0.33]{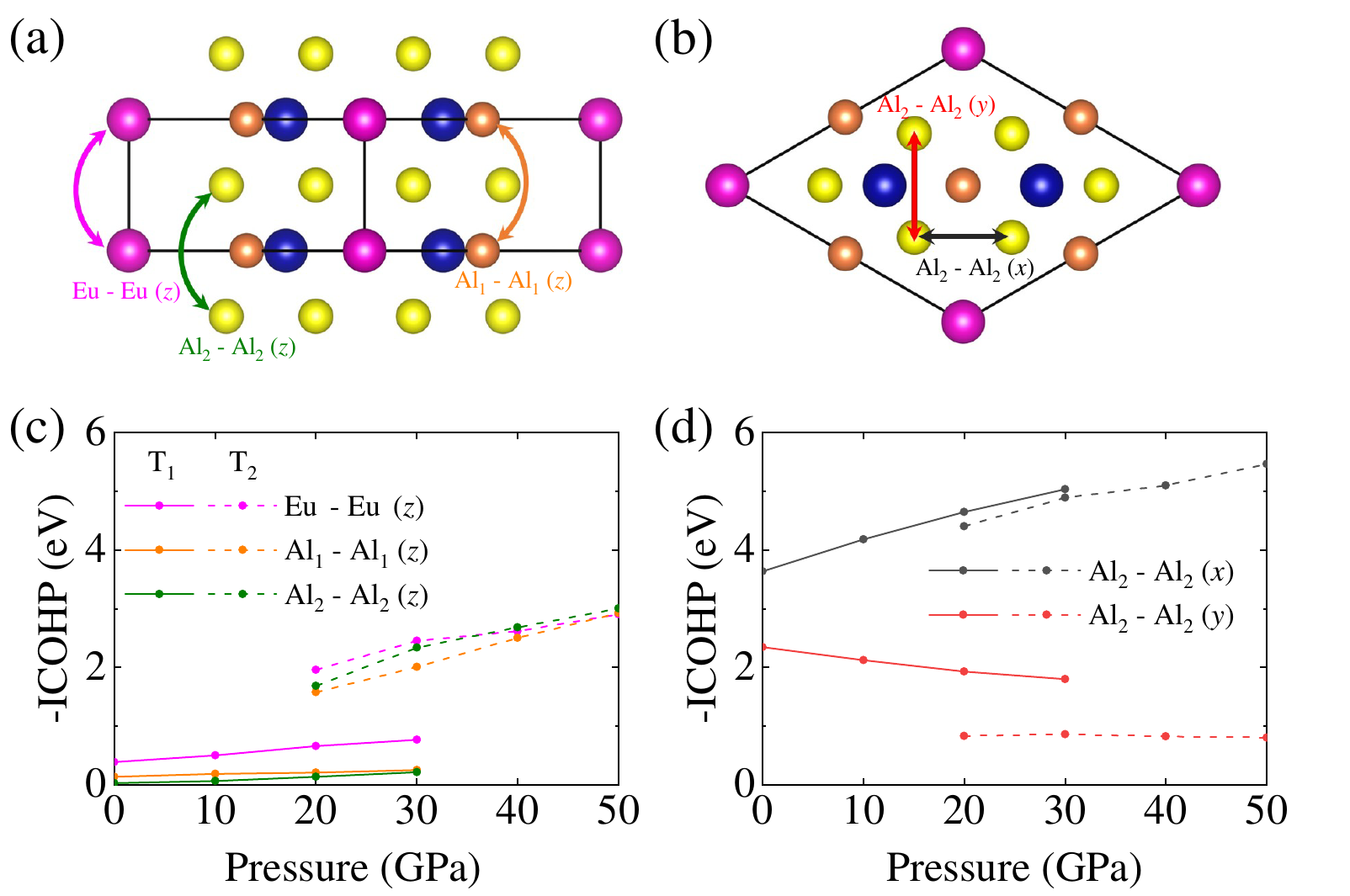}
\caption{
(a,b) Bond pairs whose bonding characteristics change most strongly under pressure. (c,d) MLWF-based $-\mathrm{ICOHP}$ values in the T$_1$ (solid curves) and T$_2$ (dashed curves) phases.
 }
\label{Fig_icohp}
\end{figure}

As discussed above, the isostructural transition in $AM_2$Al$_9$ is not accompanied by spin, charge, or other electronic symmetry breaking. 
Although there is a noticeable variation in the band structure (see Fig. S1), the symmetry of the charge distribution consistently aligns with the crystal symmetry. 
This is distinct from the strongly correlated 4f electron-driven transitions observed in the element Ce~\cite{Ce-cal,Ce-cp}. 
Because the transition features a pronounced collapse of $c_0$, it is natural to examine changes in chemical bonding. Bond-controlled isostructural transitions have also been discussed in elemental Ca~\cite{Ca} and in the Sn$_4$P$_3$ superconductor~\cite{Sn4P3}. We therefore combined MLWF construction with ICOHP analysis to determine how pressure modifies the bonds in $AM_2$Al$_9$. 

Figures \ref{Fig_icohp}(a) and (b) identify the bonds whose strengths change most strongly with pressure. 
The remaining bonds are either weak or exhibit only minor strengthening due to the intrinsic effects of pressure. 
Taking the EuCo$_2$Al$_9$ system as an example, the bonds that are sensitive to pressure include: (a) inter-layer Eu-Eu, Al$_1$-Al$_1$, and Al$_2$-Al$_2$ bonds in the \textit{z} direction, and (b) intra-plane Al$_2$-Al$_2$ bonds in the \textit{x} and \textit{y} directions. 
The interlayer bond lengths directly control $c_0$, whereas the in-plane bonds strongly influence $a_0$ and $b_0$. Panels (c) and (d) show the pressure dependence of $-\mathrm{ICOHP}$ in the T$_1$ (solid curves) and T$_2$ (dashed curves) phases. 
As illustrated in (c) and (d), before the T$_1$-T$_2$ transition, nearly all studied bonds of (a) and (b) strengthen with increased pressure except for the Al$_2$-Al$_2$(\textit{y}) bond (indicated by the red line), showing a decreasing -ICOHP with pressure. 

The anomalous pressure-induced weakening of the Al$_2$--Al$_2$($y$) bond merits attention. 
As described by eq. \ref{eq_icohp}, the ICOHP for bonding atoms is calculated by summing the products of their orbital hopping, $H_{A\alpha,B\beta}(\textbf{R})$, and bond orders, $D_{A\alpha,B\beta}(\textbf{R})$. 
Generally, under pressure, the reduction of bond length will increase the orbital overlap and hopping integral $H_{A\alpha,B\beta}(\textbf{R})$, thereby enhancing the bond strength. 
For the Al$_2$-Al$_2$(\textit{y}) bond in the T$_1$ phase, although the orbital hopping increases with pressure, the bond order $D_{A\alpha,B\beta}(\textbf{R})$ decreases. 
Consequently, its -ICOHP also decays with pressure. 
This reduction in bond order is due to the loss of electrons from its bonding states. 
The depleted charge is transferred primarily into bonding states associated with the interlayer bonds shown in panel (a). 
Specifically, the Eu-Eu(\textit{z}) bond exhibits the most significant increase in orbital filling, which is evident from the expanded electron-type Fermi pocket (see the projected orbital weight in Fig. S1(b)). 

The pressure-induced transition can thus be understood as a competition for bonding electrons between the in-plane Al$_2$--Al$_2$($y$) bond and the interlayer bonds. 
At ambient pressure, more electrons populate the bonding states of the intra-planar Al$_2$-Al$_2$(\textit{y}) bond, as shown in Fig. \ref{Fig_icohp}(b), which supports its stronger bond strength. 
This stronger bond naturally constrains the intra-planar \textbf{\textit{a$_0$}} and \textbf{\textit{b$_0$}} lattice constants, stabilizing the T$_1$ phase. 
Under pressure, electrons are transferred from the Al$_2$--Al$_2$($y$) bond into interlayer bonding states. 
Consequently, the in-plane Al$_2$--Al$_2$($y$) bond weakens, while the interlayer Eu--Eu, Al$_1$--Al$_1$, and Al$_2$--Al$_2$ bonds acquire stronger metallic character. The enhanced interlayer cohesion stabilizes the T$_2$ phase, which has a shorter $c_0$ and larger $a_0$ and $b_0$, consistent with the enhanced $-\mathrm{ICOHP}$ values in Fig. \ref{Fig_icohp}(c). Thus, the transition results from competition for bonding charge between the in-plane Al$_2$--Al$_2$($y$) bond and the interlayer $A$--$A$($z$), Al$_1$--Al$_1$($z$), and Al$_2$--Al$_2$($z$) bonds. 
Specifically, in systems with $M$ = Fe, the partially-filled Fe 3d orbitals can also capture the bonding electrons from their neighboring Al$_2$-Al$_2$(\textit{y}) bonds, potentially catalyzing the transition to the T$_2$ phase at lower pressure.

\section{Summary}

Structural phase transitions in crystals can be grouped into three broad classes: symmetry-changing first-order transitions, symmetry-changing second-order transitions, and symmetry-preserving isostructural transitions. The last class has a distinctive phase diagram in which the first-order boundary can terminate at a critical point. Using first-principles calculations, we have shown that $AM_2$Al$_9$ compounds ($A$ = Ba, Ca, Sr, or Eu; $M$ = Fe, Co, or Ni) undergo pressure-induced isostructural transitions. Across the transition, $a_0$ and $b_0$ expand abruptly, $c_0$ collapses, and the structure changes from T$_1$ to T$_2$ while retaining space group $P6/mmm$. The small calculated barrier near the transition pressure suggests that the associated critical point may be experimentally accessible. Electronic-structure and bonding analyses further show that the transition is driven primarily by pressure-induced charge transfer and competition between in-plane Al$_2$--Al$_2$ bonding and interlayer bonding. Because this combination of preserved symmetry, substantial volume collapse, and low transition barrier is unusual, $AM_2$Al$_9$ offers a promising platform for investigating critical phenomena in single-crystalline solids under pressure.

\begin{acknowledgments}
This work was supported by funding administered through HPSTAR, the National Natural Science Foundation of China (Grant No.~12204410), and the Natural Science Foundation of Zhejiang Province (Grant No.~LMS25A040002).
\end{acknowledgments}

\bibliography{129}

@article{Ca,
  title = {Topological Catastrophe and Isostructural Phase Transition in Calcium},
  author = {Jones, Travis E. and Eberhart, Mark E. and Clougherty, Dennis P.},
  journal = {Phys. Rev. Lett.},
  volume = {105},
  issue = {26},
  pages = {265702},
  numpages = {4},
  year = {2010},
  month = {Dec},
  publisher = {American Physical Society},
  doi = {10.1103/PhysRevLett.105.265702},
  url = {https://link.aps.org/doi/10.1103/PhysRevLett.105.265702}
}

@article{Fe,
    author = {Kong, L. T. and Liu, B. X.},
    title = {Correlation of magnetic moment versus spacing distance of metastable fcc structured iron},
    journal = {Applied Physics Letters},
    volume = {84},
    number = {18},
    pages = {3627-3629},
    year = {2004},
    month = {05},
    issn = {0003-6951},
    doi = {10.1063/1.1738516},
    url = {https://doi.org/10.1063/1.1738516},
}

@article{BFO-PTO,
  title={Temperature-induced isostructural phase transition, associated large negative volume expansion, and the existence of a critical point in the phase diagram of the multiferroic (1--x) BiFeO 3--x PbTiO 3 solid solution system},
  author={Bhattacharjee, Shuvrajyoti and Taji, Kazuaki and Moriyoshi, Chikako and Kuroiwa, Yoshihiro and Pandey, Dhananjai},
  journal={Physical Review B—Condensed Matter and Materials Physics},
  volume={84},
  number={10},
  pages={104116},
  year={2011},
  publisher={APS}
}

@article{VO2,
  title = {${\mathrm{VO}}_{2}$ under hydrostatic pressure: Isostructural phase transition close to a critical endpoint},
  author = {Bouvier, P. and Bussmann, L. and Machon, D. and Breslavetz, I. and Garbarino, G. and Strobel, P. and Dmitriev, V.},
  journal = {Phys. Rev. B},
  volume = {108},
  issue = {14},
  pages = {144106},
  numpages = {14},
  year = {2023},
  month = {Oct},
  publisher = {American Physical Society},
  doi = {10.1103/PhysRevB.108.144106},
  url = {https://link.aps.org/doi/10.1103/PhysRevB.108.144106}
}

@article{Zr,
  title={Anharmonicity-induced first-order isostructural phase transition of zirconium under pressure},
  author={Stavrou, Elissaios and Yang, Lin H and S{\"o}derlind, Per and Aberg, Daniel and Radousky, Harry B and Armstrong, Michael R and Belof, Jonathan L and Kunz, Martin and Greenberg, Eran and Prakapenka, Vitali B and others},
  journal={Physical Review B},
  volume={98},
  number={22},
  pages={220101},
  year={2018},
  publisher={APS}
}

@article{Pr,
  title = {High-temperature fcc phase of Pr:  Negative thermal expansion and intermediate valence state},
  author = {Kuznetsov, A. Yu. and Dmitriev, V. P. and Bandilet, O. I. and Weber, H.-P.},
  journal = {Phys. Rev. B},
  volume = {68},
  issue = {6},
  pages = {064109},
  numpages = {6},
  year = {2003},
  month = {Aug},
  publisher = {American Physical Society},
  doi = {10.1103/PhysRevB.68.064109},
  url = {https://link.aps.org/doi/10.1103/PhysRevB.68.064109}
}

@article{Ce,
  title = {Concerning the High Pressure Allotropic Modification of Cerium},
  author = {Lawson, A. W. and Tang, Ting-Yuan},
  journal = {Phys. Rev.},
  volume = {76},
  issue = {2},
  pages = {301--302},
  numpages = {0},
  year = {1949},
  month = {Jul},
  publisher = {American Physical Society},
  doi = {10.1103/PhysRev.76.301},
  url = {https://link.aps.org/doi/10.1103/PhysRev.76.301}
}

@article{Ce-cal,
  title={$\gamma$-$\alpha$ isostructural transition in cerium},
  author={Lanata, Nicola and Yao, Yong-Xin and Wang, Cai-Zhuang and Ho, Kai-Ming and Schmalian, J{\"o}rg and Haule, Kristjan and Kotliar, Gabriel},
  journal={Physical review letters},
  volume={111},
  number={19},
  pages={196801},
  year={2013},
  publisher={APS}
}

@article{Ce-cp,
title = {Martensitic-like microstructures across the isostructural phase transitions in Cerium},
journal = {Acta Materialia},
volume = {271},
pages = {119863},
year = {2024},
issn = {1359-6454},
doi = {https://doi.org/10.1016/j.actamat.2024.119863},
url = {https://www.sciencedirect.com/science/article/pii/S1359645424002167},
author = {Laura Henry and Nicolas Bruzy and Robin Fréville and Christophe Denoual and Bernard Amadon and Églantine Boulard and Andrew King and Nicolas Guignot and Agnès Dewaele}
}

@Article{WSe2,
author={Wang, Xuefei
and Chen, Xuliang
and Zhou, Yonghui
and Park, Changyong
and An, Chao
and Zhou, Ying
and Zhang, Ranran
and Gu, Chuanchuan
and Yang, Wenge
and Yang, Zhaorong},
title={Pressure-induced iso-structural phase transition and metallization in WSe2},
journal={Scientific Reports},
year={2017},
month={May},
day={04},
volume={7},
number={1},
pages={46694},
issn={2045-2322},
doi={10.1038/srep46694},
url={https://doi.org/10.1038/srep46694}
}

@Article{A2Ir2O7,
author={Rosalin, M.
and Kisku, Sebabrata
and Telang, Prachi
and Singh, Surjeet
and Muthu, D. V. S.
and Sood, A. K.},
title={Pressure-induced iso-structural phase transitions in pyrochlore iridates},
journal={Pramana},
year={2024},
month={Nov},
day={23},
volume={98},
number={4},
pages={165},
issn={0973-7111},
doi={10.1007/s12043-024-02852-w},
url={https://doi.org/10.1007/s12043-024-02852-w}
}

@article{SmS,
title = {Critical point for the S - M transition in SmS},
journal = {Solid State Communications},
volume = {26},
number = {3},
pages = {173-175},
year = {1978},
issn = {0038-1098},
doi = {https://doi.org/10.1016/0038-1098(78)91064-5},
url = {https://www.sciencedirect.com/science/article/pii/0038109878910645},
author = {V. Shubha and T.G. Ramesh and S. Ramaseshan}
}

@article{TiS3,
  title = {Breakdown of electron lone pair and insulator-to-metal transition in $\mathrm{Ti}{\mathrm{S}}_{3}$ under high pressure},
  author = {Ulla, A. R. Atique and Modak, P. and Verma, Ashok K.},
  journal = {Phys. Rev. Mater.},
  volume = {8},
  issue = {12},
  pages = {124601},
  numpages = {8},
  year = {2024},
  month = {Dec},
  publisher = {American Physical Society},
  doi = {10.1103/PhysRevMaterials.8.124601},
  url = {https://link.aps.org/doi/10.1103/PhysRevMaterials.8.124601}
}

@Article{Sn4P3,
author={Wang, Qi
and Wu, Juefei
and Wang, Yiyan
and Zheng, Fanbang
and Pei, Cuiying
and Zhao, Yi
and Cao, Weizheng
and Chen, Yulin
and Xia, Tianlong
and Yan, Shichao
and Qi, Yanpeng},
title={Giant negative area compressibility in layered Sn4P3 with enhanced superconductivity},
journal={Cell Reports Physical Science},
year={2025},
month={Feb},
day={19},
publisher={Elsevier},
volume={6},
number={2},
issn={2666-3864},
doi={10.1016/j.xcrp.2025.102450},
url={https://doi.org/10.1016/j.xcrp.2025.102450}
}

@article{NbSe2,
  title = {Pressure Induced Effects on the Fermi Surface of Superconducting $2\mathrm{H}\mathrm{\text{\ensuremath{-}}}{\mathrm{NbSe}}_{2}$},
  author = {Suderow, H. and Tissen, V. G. and Brison, J. P. and Mart\'{\i}nez, J. L. and Vieira, S.},
  journal = {Phys. Rev. Lett.},
  volume = {95},
  issue = {11},
  pages = {117006},
  numpages = {4},
  year = {2005},
  month = {Sep},
  publisher = {American Physical Society},
  doi = {10.1103/PhysRevLett.95.117006},
  url = {https://link.aps.org/doi/10.1103/PhysRevLett.95.117006}
}

@Article{TaS2,
author={Nakata, Yuki
and Sugawara, Katsuaki
and Chainani, Ashish
and Oka, Hirofumi
and Bao, Changhua
and Zhou, Shaohua
and Chuang, Pei-Yu
and Cheng, Cheng-Maw
and Kawakami, Tappei
and Saruta, Yasuaki
and Fukumura, Tomoteru
and Zhou, Shuyun
and Takahashi, Takashi
and Sato, Takafumi},
title={Robust charge-density wave strengthened by electron correlations in monolayer 1T-TaSe2 and 1T-NbSe2},
journal={Nature Communications},
year={2021},
month={Oct},
day={07},
volume={12},
number={1},
pages={5873},
issn={2041-1723},
doi={10.1038/s41467-021-26105-1},
url={https://doi.org/10.1038/s41467-021-26105-1}
}

@article{PbTiO3,
title = {Ferroelectric properties of PbTiO3},
journal = {Physica},
volume = {28},
number = {9},
pages = {871-876},
year = {1962},
issn = {0031-8914},
doi = {https://doi.org/10.1016/0031-8914(62)90075-7},
url = {https://www.sciencedirect.com/science/article/pii/0031891462900757},
author = {V.G. Bhide and K.G. Deshmukh and M.S. Hegde}
}

@Article{BaTiO3,
author={Smith, Millicent B.
and Page, Katharine
and Siegrist, Theo
and Redmond, Peter L.
and Walter, Erich C.
and Seshadri, Ram
and Brus, Louis E.
and Steigerwald, Michael L.},
title={Crystal Structure and the Paraelectric-to-Ferroelectric Phase Transition of Nanoscale BaTiO3},
journal={Journal of the American Chemical Society},
year={2008},
month={Jun},
day={01},
publisher={American Chemical Society},
volume={130},
number={22},
pages={6955-6963},
issn={0002-7863},
doi={10.1021/ja0758436},
url={https://doi.org/10.1021/ja0758436}
}

@article{129-1,
title = {Zur kenntnis des BaFe2Al9-strukturtyps: Ternäre aluminide at2Al9 MIT A = Ba, Sr und T = Fe, Co, Ni},
journal = {Journal of the Less Common Metals},
volume = {40},
number = {1},
pages = {91-96},
year = {1975},
issn = {0022-5088},
doi = {https://doi.org/10.1016/0022-5088(75)90184-8},
url = {https://www.sciencedirect.com/science/article/pii/0022508875901848},
author = {K. Turban and Herbert Schäfer}
}

@Article{BaFeAl-1,
author={Meier, William R.
and Chakoumakos, Bryan C.
and Okamoto, Satoshi
and McGuire, Michael A.
and Hermann, Rapha{\"e}l P.
and Samolyuk, German D.
and Gao, Shang
and Zhang, Qiang
and Stone, Matthew B.
and Christianson, Andrew D.
and Sales, Brian C.},
title={A Catastrophic Charge Density Wave in BaFe2Al9},
journal={Chemistry of Materials},
year={2021},
month={Apr},
day={27},
publisher={American Chemical Society},
volume={33},
number={8},
pages={2855-2863},
issn={0897-4756},
doi={10.1021/acs.chemmater.1c00005},
url={https://doi.org/10.1021/acs.chemmater.1c00005}
}

@article{BaFeAl-2,
  title = {Origin of the charge density wave state in ${\mathrm{BaFe}}_{2}{\mathrm{Al}}_{9}$},
  author = {Li, Yuping and Liu, Mingfeng and Li, Jiangxu and Wang, Jiantao and Lai, Junwen and He, Dongchang and Qiu, Ruizhi and Sun, Yan and Chen, Xing-Qiu and Liu, Peitao},
  journal = {Phys. Rev. B},
  volume = {110},
  issue = {19},
  pages = {195118},
  numpages = {8},
  year = {2024},
  month = {Nov},
  publisher = {American Physical Society},
  doi = {10.1103/PhysRevB.110.195118},
  url = {https://link.aps.org/doi/10.1103/PhysRevB.110.195118}
}

@article{BaFeAl-3,
  title = {$^{27}\mathrm{Al}$ NMR insight into the phase transition in $\mathrm{Ba}{\mathrm{Fe}}_{2}{\mathrm{Al}}_{9}$},
  author = {Huang, C. Y. and Lee, H. Y. and Chang, Y. C. and Hong, Chon Kit and Ou, Y. R. and Kuo, C. N. and Lue, C. S.},
  journal = {Phys. Rev. B},
  volume = {106},
  issue = {19},
  pages = {195101},
  numpages = {6},
  year = {2022},
  month = {Nov},
  publisher = {American Physical Society},
  doi = {10.1103/PhysRevB.106.195101},
  url = {https://link.aps.org/doi/10.1103/PhysRevB.106.195101}
}

@article{BaFeAl-4,
  title = {Thermal characteristics of the phase transition near 100 K in ${\mathrm{BaFe}}_{2}{\mathrm{Al}}_{9}$},
  author = {Kuo, C. N. and Huang, R. Y. and Wen, L. T. and Lee, H. Y. and Hong, C. K. and Ou, Y. R. and Kuo, Y. K. and Lue, C. S.},
  journal = {Phys. Rev. B},
  volume = {110},
  issue = {4},
  pages = {045128},
  numpages = {6},
  year = {2024},
  month = {Jul},
  publisher = {American Physical Society},
  doi = {10.1103/PhysRevB.110.045128},
  url = {https://link.aps.org/doi/10.1103/PhysRevB.110.045128}
}

@article{BaCoAl,
  title = {Electronic structure of intertwined kagome, honeycomb, and triangular sublattices of the intermetallics $M{\mathrm{Co}}_{2}{\mathrm{Al}}_{9}$ ($M$ = Sr, Ba)},
  author = {Bigi, Chiara and Pakdel, Sahar and Winiarski, Micha\l{} J. and Orgiani, Pasquale and Vobornik, Ivana and Fujii, Jun and Rossi, Giorgio and Polewczyk, Vincent and King, Phil D. C. and Panaccione, Giancarlo and Klimczuk, Tomasz and Thygesen, Kristian Sommer and Mazzola, Federico},
  journal = {Phys. Rev. B},
  volume = {108},
  issue = {7},
  pages = {075148},
  numpages = {7},
  year = {2023},
  month = {Aug},
  publisher = {American Physical Society},
  doi = {10.1103/PhysRevB.108.075148},
  url = {https://link.aps.org/doi/10.1103/PhysRevB.108.075148}
}

@article{EuCoAl,
  author = {V. M. T. Thiede and W. Jeitschko},
  title = {Crystal Structure of Europium Cobalt Aluminide (1/2/9), EuCo2Al9},
  journal = {Zeitschrift f{\"u}r Kristallographie -- New Crystal Structures},
  volume = {214},
  number = {2},
  pages = {149--150},
  year = {1999},
  doi = {10.1515/ncrs-1999-0205},
  url = {https://doi.org/10.1515/ncrs-1999-0205}
}

@Article{129-2,
author={Calta, Nicholas P.
and Han, Fei
and Kanatzidis, Mercouri G.},
title={Synthesis, Structure, and Rigid Unit Mode-like Anisotropic Thermal Expansion of BaIr2In9},
journal={Inorganic Chemistry},
year={2015},
month={Sep},
day={08},
publisher={American Chemical Society},
volume={54},
number={17},
pages={8794-8799},
issn={0020-1669},
doi={10.1021/acs.inorgchem.5b01421},
url={https://doi.org/10.1021/acs.inorgchem.5b01421}
}

@Article{129-3,
author={Lei, Xiao-Wu
and Zhong, Guo-Hua
and Li, Long-Hua
and Hu, Chun-Li
and Li, Min-Jie
and Mao, Jiang-Gao},
title={Eu3Co2In15 and KM2In9 (M = Co, Ni): 3D Frameworks Based on Transition Metal Centered In9 Clusters},
journal={Inorganic Chemistry},
year={2009},
month={Mar},
day={16},
publisher={American Chemical Society},
volume={48},
number={6},
pages={2526-2533},
issn={0020-1669},
doi={10.1021/ic8019765},
url={https://doi.org/10.1021/ic8019765}
}

@article{129-4,
title = {Single crystal growth and physical properties of MCo2Al9 (M= Sr, Ba)},
journal = {Journal of Solid State Chemistry},
volume = {289},
pages = {121509},
year = {2020},
issn = {0022-4596},
doi = {https://doi.org/10.1016/j.jssc.2020.121509},
url = {https://www.sciencedirect.com/science/article/pii/S002245962030339X},
author = {Zuzanna Ryżyńska and Tomasz Klimczuk and Michał J. Winiarski}
}

@article{1982,
  title={On the theory of isostructural phase transitions in crystals},
  author={Schneider, VE and Tornau, EE},
  journal={physica status solidi (b)},
  volume={111},
  number={2},
  pages={565--574},
  year={1982},
  publisher={Wiley Online Library}
}

@article{1989,
  title={Theory of isostructural phase transitions described by two order parameters},
  author={Bezrukov, GV and Men, AN and Talanov, VM},
  journal={physica status solidi (a)},
  volume={116},
  number={2},
  pages={603--613},
  year={1989},
  publisher={WILEY-VCH Verlag Berlin}
}

@article{dft1,
  title = {Inhomogeneous Electron Gas},
  author = {Hohenberg, P. and Kohn, W.},
  journal = {Phys. Rev.},
  volume = {136},
  issue = {3B},
  pages = {B864--B871},
  numpages = {0},
  year = {1964},
  month = {Nov},
  publisher = {American Physical Society},
  doi = {10.1103/PhysRev.136.B864},
  url = {https://link.aps.org/doi/10.1103/PhysRev.136.B864}
}

@article{dft2,
  title = {Self-Consistent Equations Including Exchange and Correlation Effects},
  author = {Kohn, W. and Sham, L. J.},
  journal = {Phys. Rev.},
  volume = {140},
  issue = {4A},
  pages = {A1133--A1138},
  numpages = {0},
  year = {1965},
  month = {Nov},
  publisher = {American Physical Society},
  doi = {10.1103/PhysRev.140.A1133},
  url = {https://link.aps.org/doi/10.1103/PhysRev.140.A1133}
}

@article{pbe,
  title = {Generalized Gradient Approximation Made Simple},
  author = {Perdew, John P. and Burke, Kieron and Ernzerhof, Matthias},
  journal = {Phys. Rev. Lett.},
  volume = {77},
  issue = {18},
  pages = {3865--3868},
  numpages = {0},
  year = {1996},
  month = {Oct},
  publisher = {American Physical Society},
  doi = {10.1103/PhysRevLett.77.3865},
  url = {https://link.aps.org/doi/10.1103/PhysRevLett.77.3865}
}

@article{dfptreview,
  title = {Electron-phonon interactions from first principles},
  author = {Giustino, Feliciano},
  journal = {Rev. Mod. Phys.},
  volume = {89},
  issue = {1},
  pages = {015003},
  numpages = {63},
  year = {2017},
  month = {Feb},
  publisher = {American Physical Society},
  doi = {10.1103/RevModPhys.89.015003},
  url = {https://link.aps.org/doi/10.1103/RevModPhys.89.015003}
}

@article{dfptreview2,
  title = {Phonons and related crystal properties from density-functional perturbation theory},
  author = {Baroni, Stefano and de Gironcoli, Stefano and Dal Corso, Andrea and Giannozzi, Paolo},
  journal = {Rev. Mod. Phys.},
  volume = {73},
  issue = {2},
  pages = {515--562},
  numpages = {0},
  year = {2001},
  month = {Jul},
  publisher = {American Physical Society},
  doi = {10.1103/RevModPhys.73.515},
  url = {https://link.aps.org/doi/10.1103/RevModPhys.73.515}
}

@article{pwscf,
	doi = {10.1088/0953-8984/21/39/395502},
	url = {https://doi.org/10.1088/0953-8984/21/39/395502},
	year = 2009,
	month = {sep},
	publisher = {{IOP} Publishing},
	volume = {21},
	number = {39},
	pages = {395502},
	author = {Paolo Giannozzi and Stefano Baroni and Nicola Bonini and Matteo Calandra and Roberto Car and Carlo Cavazzoni and Davide Ceresoli and Guido L Chiarotti and Matteo Cococcioni and Ismaila Dabo and Andrea Dal Corso and Stefano de Gironcoli and Stefano Fabris and Guido Fratesi and Ralph Gebauer and Uwe Gerstmann and Christos Gougoussis and Anton Kokalj and Michele Lazzeri and Layla Martin-Samos and Nicola Marzari and Francesco Mauri and Riccardo Mazzarello and Stefano Paolini and Alfredo Pasquarello and Lorenzo Paulatto and Carlo Sbraccia and Sandro Scandolo and Gabriele Sclauzero and Ari P Seitsonen and Alexander Smogunov and Paolo Umari and Renata M Wentzcovitch},
	title = {{QUANTUM} {ESPRESSO}: a modular and open-source software project for quantum simulations of materials},
	journal = {J. Phys.: Condens. Matter},
}

@article{mlwf,
title = {An updated version of wannier90: A tool for obtaining maximally-localised Wannier functions},
journal = {Comput. Phys. Commun.},
volume = {185},
number = {8},
pages = {2309-2310},
year = {2014},
issn = {0010-4655},
doi = {https://doi.org/10.1016/j.cpc.2014.05.003},
url = {https://www.sciencedirect.com/science/article/pii/S001046551400157X},
author = {Arash A. Mostofi and Jonathan R. Yates and Giovanni Pizzi and Young-Su Lee and Ivo Souza and David Vanderbilt and Nicola Marzari},

}

@article{cohp,
author = {Deringer, Volker L. and Tchougréeff, Andrei L. and Dronskowski, Richard},
title = {Crystal Orbital Hamilton Population (COHP) Analysis As Projected from Plane-Wave Basis Sets},
journal = {J. Phys. Chem. A},
volume = {115},
number = {21},
pages = {5461-5466},
year = {2011},
doi = {10.1021/jp202489s},
URL = { https://doi.org/10.1021/jp202489s}
    
}

@article{cohp2,
author = {Maintz, Stefan and Deringer, Volker L. and Tchougréeff, Andrei L. and Dronskowski, Richard},
title = {Analytic projection from plane-wave and PAW wavefunctions and application to chemical-bonding analysis in solids},
journal = {J. Comput. Chem.},
volume = {34},
number = {29},
pages = {2557-2567},
doi = {https://doi.org/10.1002/jcc.23424},
url = {https://onlinelibrary.wiley.com/doi/abs/10.1002/jcc.23424},
year = {2013}
}

@article{vasp1,
  title = {Ab initio molecular dynamics for liquid metals},
  author = {Kresse, G. and Hafner, J.},
  journal = {Phys. Rev. B},
  volume = {47},
  issue = {1},
  pages = {558--561},
  numpages = {0},
  year = {1993},
  month = {Jan},
  publisher = {American Physical Society},
  doi = {10.1103/PhysRevB.47.558},
  url = {https://link.aps.org/doi/10.1103/PhysRevB.47.558}
}

@article{vasp2,
  title = {Efficient iterative schemes for ab initio total-energy calculations using a plane-wave basis set},
  author = {Kresse, G. and Furthm\"uller, J.},
  journal = {Phys. Rev. B},
  volume = {54},
  issue = {16},
  pages = {11169--11186},
  numpages = {0},
  year = {1996},
  month = {Oct},
  publisher = {American Physical Society},
  doi = {10.1103/PhysRevB.54.11169},
  url = {https://link.aps.org/doi/10.1103/PhysRevB.54.11169}
}

@article{ldau,
  title = {Linear response approach to the calculation of the effective interaction parameters in the $\mathrm{LDA}+\mathrm{U}$ method},
  author = {Cococcioni, Matteo and de Gironcoli, Stefano},
  journal = {Phys. Rev. B},
  volume = {71},
  issue = {3},
  pages = {035105},
  numpages = {16},
  year = {2005},
  month = {Jan},
  publisher = {American Physical Society},
  doi = {10.1103/PhysRevB.71.035105},
  url = {https://link.aps.org/doi/10.1103/PhysRevB.71.035105}
}

@Article{NbSe2-cal,
author={Lian, Chao-Sheng
and Si, Chen
and Duan, Wenhui},
title={Unveiling Charge-Density Wave, Superconductivity, and Their Competitive Nature in Two-Dimensional NbSe2},
journal={Nano Letters},
year={2018},
month={May},
day={09},
publisher={American Chemical Society},
volume={18},
number={5},
pages={2924-2929},
issn={1530-6984},
doi={10.1021/acs.nanolett.8b00237},
url={https://doi.org/10.1021/acs.nanolett.8b00237}
}

@article{TaS2-cal,
title = {Modulating charge density wave states in 1T-TaS2 by self-intercalation: A DFT study},
journal = {Materials Today Communications},
volume = {40},
pages = {109388},
year = {2024},
issn = {2352-4928},
doi = {https://doi.org/10.1016/j.mtcomm.2024.109388},
url = {https://www.sciencedirect.com/science/article/pii/S2352492824013692},
author = {Jia Wei and Jiming Zheng and Min Wang and Guoguo Tian and Sujuan Zhang and Guo Ping}
}

@Article{TiSe2-cal,
author={Yin, Li
and Tang, Hong
and Berlijn, Tom
and Ruzsinszky, Adrienn},
title={Efficient simulations of charge density waves in the transition metal Dichalcogenide TiSe2},
journal={npj Computational Materials},
year={2024},
month={Sep},
day={07},
volume={10},
number={1},
pages={207},
issn={2057-3960},
doi={10.1038/s41524-024-01396-2},
url={https://doi.org/10.1038/s41524-024-01396-2}
}

@article{CsV3Sb5-cal,
  title = {Charge Density Waves and Electronic Properties of Superconducting Kagome Metals},
  author = {Tan, Hengxin and Liu, Yizhou and Wang, Ziqiang and Yan, Binghai},
  journal = {Phys. Rev. Lett.},
  volume = {127},
  issue = {4},
  pages = {046401},
  numpages = {6},
  year = {2021},
  month = {Jul},
  publisher = {American Physical Society},
  doi = {10.1103/PhysRevLett.127.046401},
  url = {https://link.aps.org/doi/10.1103/PhysRevLett.127.046401}
}

@article{BaTiO3-cal,
title = {BaTiO3: Energy, geometrical and electronic structure, relationship between optical constant and density from first-principles calculations},
journal = {Optical Materials},
volume = {35},
number = {12},
pages = {2629-2637},
year = {2013},
issn = {0925-3467},
doi = {https://doi.org/10.1016/j.optmat.2013.07.034},
url = {https://www.sciencedirect.com/science/article/pii/S0925346713004163},
author = {Qi-Jun Liu and Ning-Chao Zhang and Fu-Sheng Liu and Hong-Yan Wang and Zheng-Tang Liu}
}

@article{PbTiO3-cal,
title = {First-principle calculations of the cohesive energy and the electronic properties of PbTiO3},
journal = {Physica B: Condensed Matter},
volume = {391},
number = {2},
pages = {316-321},
year = {2007},
issn = {0921-4526},
doi = {https://doi.org/10.1016/j.physb.2006.10.013},
url = {https://www.sciencedirect.com/science/article/pii/S0921452606017182},
author = {S.M. Hosseini and T. Movlarooy and A. Kompany}
}

@Article{PbZrO3-cal,
author={Tagantsev, A. K.
and Vaideeswaran, K.
and Vakhrushev, S. B.
and Filimonov, A. V.
and Burkovsky, R. G.
and Shaganov, A.
and Andronikova, D.
and Rudskoy, A. I.
and Baron, A. Q. R.
and Uchiyama, H.
and Chernyshov, D.
and Bosak, A.
and Ujma, Z.
and Roleder, K.
and Majchrowski, A.
and Ko, J.-H.
and Setter, N.},
title={The origin of antiferroelectricity in PbZrO3},
journal={Nature Communications},
year={2013},
month={Jul},
day={29},
volume={4},
number={1},
pages={2229},
issn={2041-1723},
doi={10.1038/ncomms3229},
url={https://doi.org/10.1038/ncomms3229}
}

@article{CuInP2S6-cal,
author = {Wei, Xian-Kui and Domingo, Neus and Sun, Young and Balke, Nina and Dunin-Borkowski, Rafal E. and Mayer, Joachim},
title = {Progress on Emerging Ferroelectric Materials for Energy Harvesting, Storage and Conversion},
journal = {Advanced Energy Materials},
volume = {12},
number = {24},
pages = {2201199},
doi = {https://doi.org/10.1002/aenm.202201199},
year = {2022}
}

@article{xu,
title = {Giant anomalous Hall conductivity in frustrated magnet EuCo2Al9},
journal = {Materials Today},
volume = {95},
pages = {103285},
year = {2026},
issn = {1369-7021},
doi = {https://doi.org/10.1016/j.mattod.2026.103285},
url = {https://www.sciencedirect.com/science/article/pii/S1369702126001318},
author = {Sheng Xu and Jian-Feng Zhang and Shu-Xiang Li and Junfa Lin and Xiaobai Ma and Wenyun Yang and Jun-Jian Mi and Zheng Li and Tian-Hao Li and Yue-Yang Wu and Jiang Ma and Qian Tao and Wen-He Jiao and Xiaofeng Xu and Zengwei Zhu and Yuanfeng Xu and Hanjie Guo and Tian-Long Xia and Zhu-An Xu}
}

\end{document}